\documentclass[aps,prl,twocolumn,superscriptaddress]{revtex4-1}

\usepackage{amsmath}
\usepackage{amsfonts}
\usepackage{amssymb}
\usepackage{lineno}
\usepackage{booktabs}
\usepackage{multirow}
\usepackage[T1]{fontenc}

\usepackage{float}
\usepackage{graphicx}
\usepackage[caption=false]{subfig}
\usepackage{lipsum} 

\usepackage[colorlinks]{hyperref}
\newcommand{\pd}[1] {$p$-$d${\bf #1}}

\newcommand{\dd}[1] {$d$-$d${\bf #1}}
\newcommand{\pp}[1] {$p$-$p${\bf #1}}
\newcommand{\app}[1] {$\bar{p}$-$\bar{p}${\bf #1}}
\newcommand{\pphi}[1] {$p$-$\phi${\bf #1}}
\newcommand{\kp}[1] {$K^{-}$-$p${\bf #1}}
\newcommand{\pla}[1] {$p$-$\Lambda${\bf #1}}
\newcommand{\pXi}[1] {$p$-$\Xi${\bf #1}}
\newcommand{\pOmega}[1] {$p$-$\Omega${\bf #1}}
\newcommand{\lala}[1] {$\Lambda$-$\Lambda${\bf #1}}
\newcommand{\fzero}[1] {$f_0${\bf #1}}
\newcommand{\dzero}[1] {$d_0${\bf #1}}

\newcommand{\sNN}{$\sqrt{s_{\mathrm{NN}}}$}

\usepackage[utf8]{inputenc}

\begin{document}
\title{Light Nuclei Femtoscopy and Baryon Interactions in 3 GeV Au+Au Collisions at RHIC}

\affiliation{Academia Sinica}
\affiliation{Abilene Christian University, Abilene, Texas   79699}
\affiliation{AGH University of Krakow, FPACS, Cracow 30-059, Poland}
\affiliation{Argonne National Laboratory, Argonne, Illinois 60439}
\affiliation{American University in Cairo, New Cairo 11835, Egypt}
\affiliation{Ball State University, Muncie, Indiana, 47306}
\affiliation{Brookhaven National Laboratory, Upton, New York 11973}
\affiliation{University of Calabria \& INFN-Cosenza, Rende 87036, Italy}
\affiliation{University of California, Berkeley, California 94720}
\affiliation{University of California, Davis, California 95616}
\affiliation{University of California, Los Angeles, California 90095}
\affiliation{University of California, Riverside, California 92521}
\affiliation{Central China Normal University, Wuhan, Hubei 430079 }
\affiliation{University of Illinois at Chicago, Chicago, Illinois 60607}
\affiliation{Chongqing University, Chongqing, 401331}
\affiliation{Creighton University, Omaha, Nebraska 68178}
\affiliation{Czech Technical University in Prague, FNSPE, Prague 115 19, Czech Republic}
\affiliation{Technische Universit\"at Darmstadt, Darmstadt 64289, Germany}
\affiliation{National Institute of Technology Durgapur, Durgapur - 713209, India}
\affiliation{ELTE E\"otv\"os Lor\'and University, Budapest, Hungary H-1117}
\affiliation{Frankfurt Institute for Advanced Studies FIAS, Frankfurt 60438, Germany}
\affiliation{Fudan University, Shanghai, 200433 }
\affiliation{Guangxi Normal University, Guilin, 541004}
\affiliation{University of Heidelberg, Heidelberg 69120, Germany }
\affiliation{University of Houston, Houston, Texas 77204}
\affiliation{Huzhou University, Huzhou, Zhejiang  313000}
\affiliation{Indian Institute of Science Education and Research (IISER), Berhampur 760010 , India}
\affiliation{Indian Institute of Science Education and Research (IISER) Tirupati, Tirupati 517507, India}
\affiliation{Indian Institute Technology, Patna, Bihar 801106, India}
\affiliation{Indiana University, Bloomington, Indiana 47408}
\affiliation{Institute of Modern Physics, Chinese Academy of Sciences, Lanzhou, Gansu 730000 }
\affiliation{University of Jammu, Jammu 180001, India}
\affiliation{Kent State University, Kent, Ohio 44242}
\affiliation{University of Kentucky, Lexington, Kentucky 40506-0055}
\affiliation{Lanzhou University}
\affiliation{Lawrence Berkeley National Laboratory, Berkeley, California 94720}
\affiliation{Lehigh University, Bethlehem, Pennsylvania 18015}
\affiliation{Max-Planck-Institut f\"ur Physik, Munich 80805, Germany}
\affiliation{Michigan State University, East Lansing, Michigan 48824}
\affiliation{National Institute of Science Education and Research, HBNI, Jatni 752050, India}
\affiliation{National Cheng Kung University, Tainan 70101 }
\affiliation{Nuclear Physics Institute of the CAS, Rez 250 68, Czech Republic}
\affiliation{The Ohio State University, Columbus, Ohio 43210}
\affiliation{Panjab University, Chandigarh 160014, India}
\affiliation{Purdue University, West Lafayette, Indiana 47907}
\affiliation{Rice University, Houston, Texas 77251}
\affiliation{Rutgers University, Piscataway, New Jersey 08854}
\affiliation{University of Science and Technology of China, Hefei, Anhui 230026}
\affiliation{South China Normal University, Guangzhou, Guangdong 510631}
\affiliation{Sejong University, Seoul, 05006, South Korea}
\affiliation{Shandong University, Qingdao, Shandong 266237}
\affiliation{Shanghai Institute of Applied Physics, Chinese Academy of Sciences, Shanghai 201800}
\affiliation{Southern Connecticut State University, New Haven, Connecticut 06515}
\affiliation{State University of New York, Stony Brook, New York 11794}
\affiliation{Instituto de Alta Investigaci\'on, Universidad de Tarapac\'a, Arica 1000000, Chile}
\affiliation{Temple University, Philadelphia, Pennsylvania 19122}
\affiliation{Texas A\&M University, College Station, Texas 77843}
\affiliation{University of Texas, Austin, Texas 78712}
\affiliation{Tsinghua University, Beijing 100084}
\affiliation{University of Tsukuba, Tsukuba, Ibaraki 305-8571, Japan}
\affiliation{University of Chinese Academy of Sciences, Beijing, 101408}
\affiliation{United States Naval Academy, Annapolis, Maryland 21402}
\affiliation{Valparaiso University, Valparaiso, Indiana 46383}
\affiliation{Variable Energy Cyclotron Centre, Kolkata 700064, India}
\affiliation{Warsaw University of Technology, Warsaw 00-661, Poland}
\affiliation{Wayne State University, Detroit, Michigan 48201}
\affiliation{Wuhan University of Science and Technology, Wuhan, Hubei 430065}
\affiliation{Yale University, New Haven, Connecticut 06520}

\author{B.~E.~Aboona}\affiliation{Texas A\&M University, College Station, Texas 77843}
\author{J.~Adam}\affiliation{Czech Technical University in Prague, FNSPE, Prague 115 19, Czech Republic}
\author{L.~Adamczyk}\affiliation{AGH University of Krakow, FPACS, Cracow 30-059, Poland}
\author{I.~Aggarwal}\affiliation{Panjab University, Chandigarh 160014, India}
\author{M.~M.~Aggarwal}\affiliation{Panjab University, Chandigarh 160014, India}
\author{Z.~Ahammed}\affiliation{Variable Energy Cyclotron Centre, Kolkata 700064, India}
\author{E.~C.~Aschenauer}\affiliation{Brookhaven National Laboratory, Upton, New York 11973}
\author{S.~Aslam}\affiliation{Indian Institute Technology, Patna, Bihar 801106, India}
\author{J.~Atchison}\affiliation{Abilene Christian University, Abilene, Texas   79699}
\author{V.~Bairathi}\affiliation{Instituto de Alta Investigaci\'on, Universidad de Tarapac\'a, Arica 1000000, Chile}
% ...

\collaboration{STAR Collaboration}\noaffiliation

\begin{abstract} 
We report the measurements of proton-deuteron (\pd{}) and deuteron-deuteron (\dd{}) correlation functions in Au+Au collisions at $\sqrt{s_\mathrm{NN}}$ = 3\,GeV using fixed-target mode with the STAR experiment at the Relativistic Heavy-Ion Collider (RHIC). For the first time, the source size ($R_{G}$), scattering length (\fzero{}), and effective range (\dzero{}) are extracted from the measured correlation functions with a simultaneous fit. The spin-averaged $f_0$ for \pd{} and \dd{} interactions are determined to be -5.28 $\pm$ 0.11(stat.) $\pm$ 0.82(syst.) fm and -2.62 $\pm$ 0.02(stat.) $\pm$ 0.24(syst.) fm, respectively. 
The measured \pd{} interaction is consistent with theoretical calculations and low-energy scattering experiment results, demonstrating the feasibility of extracting interaction parameters using the femtoscopy technique.
The reasonable agreement between the experimental data and the calculations from the transport model indicates that deuteron production in these collisions is primarily governed by nucleon coalescence. 
\end{abstract}		
\maketitle

\section{Introduction}\label{Introduction}
In heavy-ion collisions, the measurements of two-particle femtoscopic correlations have proven to be a powerful tool for gaining insights into the space-time geometry of the particle emitting sources, as well as the interactions between pairs of particles ~\cite{Lednicky:2005af,Wang:1999bf,Chen:2024zwk,STAR:2014shf,ALICE:2020mfd}. Experimental efforts have been devoted to studying the strong interactions between particles measured in heavy-ion collisions at RHIC and LHC energies~\cite{STAR:2015kha,ALICE:2023bny,ALICE:2020mfd}. However, a majority of the studies have focused on the interactions between light and strange hadrons, such as \pp{}~\cite{ALICE:2018ysd}, \app{}~\cite{STAR:2015kha}, \pphi{}~\cite{ALICE:2021cpv}, \kp{}~\cite{ALICE:2019gcn}, \pla{}~\cite{STAR:2005rpl,ALICE:2018ysd,ALICE:2021njx}, \pXi{}~\cite{ALICE:2020mfd}, \pOmega{}~\cite{STAR:2018uho,ALICE:2020mfd}, and \lala{}~\cite{STAR:2014dcy,ALICE:2018ysd} pairs. Light nuclei, such as the deuteron, consist of loosely bound nucleons with binding energies on the order of several MeV. Conducting femtoscopic measurements between pairs of particles involving light nuclei, such as \pd{} and \dd{}, has significant relevance for the investigation of few-nucleon systems. 
These systems serve as crucial testing grounds for three-body nuclear forces, which may be essential for a more comprehensive understanding of the properties of the dense matter~\cite{Gerstung:2020ktv,Lonardoni:2014bwa,Hebeler:2015hla,ALICE:2023bny}.

The production of light nuclei in heavy-ion collisions has been extensively explored both experimentally~\cite{ALICE:2015wav,STAR:2016ydv,ALICE:2017xrp,STAR:2019sjh,STAR:2022hbp,ALICE:2022veq,STAR:2023uxk,Chen:2024zwk,E864:2000auv,E878:1998vna} and theoretically~\cite{Zhao:2022xkz,Glassel:2021rod,Staudenmaier:2021lrg,Zhao:2020irc}. The statistical thermal model~\cite{Halemane:1982ac,Andronic:2010qu} and nucleon coalescence~\cite{Csernai:1986qf,Oh:2009gx,Sato:1981ez} are the two most popular models proposed to explain their production in heavy-ion collisions. The statistical thermal model assumes that light nuclei are formed in chemical equilibrium in a thermalized system, while the nucleon coalescence model suggests that light nuclei are formed by the coalescence of nucleons at the final stage of the collision. 
It is important to note that the role of coalescence in light nuclei production, particularly at lower energies, has been discussed extensively in previous work, such as that at the STAR~\cite{STAR:2021yiu,STAR:2022hbp,STAR:2023uxk} and AGS~\cite{E864:2000auv,E878:1998vna}. Femtoscopic correlations, which provide information about the spatial distribution of particle emissions, are particularly sensitive to the size of the emitting source.  Therefore, by analyzing the femtoscopic correlations of light nuclei, we can also gain further insight into the nature of their production mechanisms.

\section{Experiment and Data Analysis}\label{The Experiment}
In this Letter, we report the femtoscopic measurement of two-particle correlation functions of light nuclei pairs, \pd{}, and \dd{}, in Au+Au collisions at $\sqrt{s_{\rm NN}}$ = 3\,GeV. The data was recorded by the Solenoidal Tracker at RHIC (STAR) under the fixed-target configuration~\cite{Meehan:2016iyt} in the year 2018. A beam of gold nuclei of energy 3.85 GeV/u was incident on a gold target of thickness 0.25 mm, corresponding to 1\% of an interaction length. The target was installed inside the vacuum pipe, 2 cm below the center of the normal beam axis, and located 200.7 cm to the west of the center of the STAR detector. 
The minimum bias (MB) trigger condition was provided by simultaneous signals from the Beam-Beam Counters (BBC) \cite{Bieser:2002ah} and the Time of Flight (TOF) detector \cite{Llope:2012zz}.
To remove collisions between the beam and beam pipe,  the reconstructed collision vertex position along the beam direction is required to be within $\pm$ 2 cm and the primary vertex position in the radial plane is required to be located within $\pm$ 1.5 cm from the center of the Au target. In total, approximately 2.6 $\times$ $10^{8}$ events pass the selection criteria and are used in this analysis.
The centrality of collisions, which quantifies the degree of nuclear overlap, is characterized by the Glauber model \cite{Ray:2007av,Miller:2007ri} fitting of the charged tracks measured in the Time Projection Chamber (TPC) \cite{Anderson:2003ur} within the pseudo-rapidity ($\eta$) region $-2 < \eta < 0$. Here, $\eta$ is defined relative to the collision vertex.  Further details about the experimental setup are given in Ref. \cite{STAR:2020dav}. The 0-60\% central events, corresponding to the 60\% of inelastic events with the smallest impact parameters, are used in this analysis.
\begin{figure}[htbp]
    \includegraphics[width=1.00\linewidth]{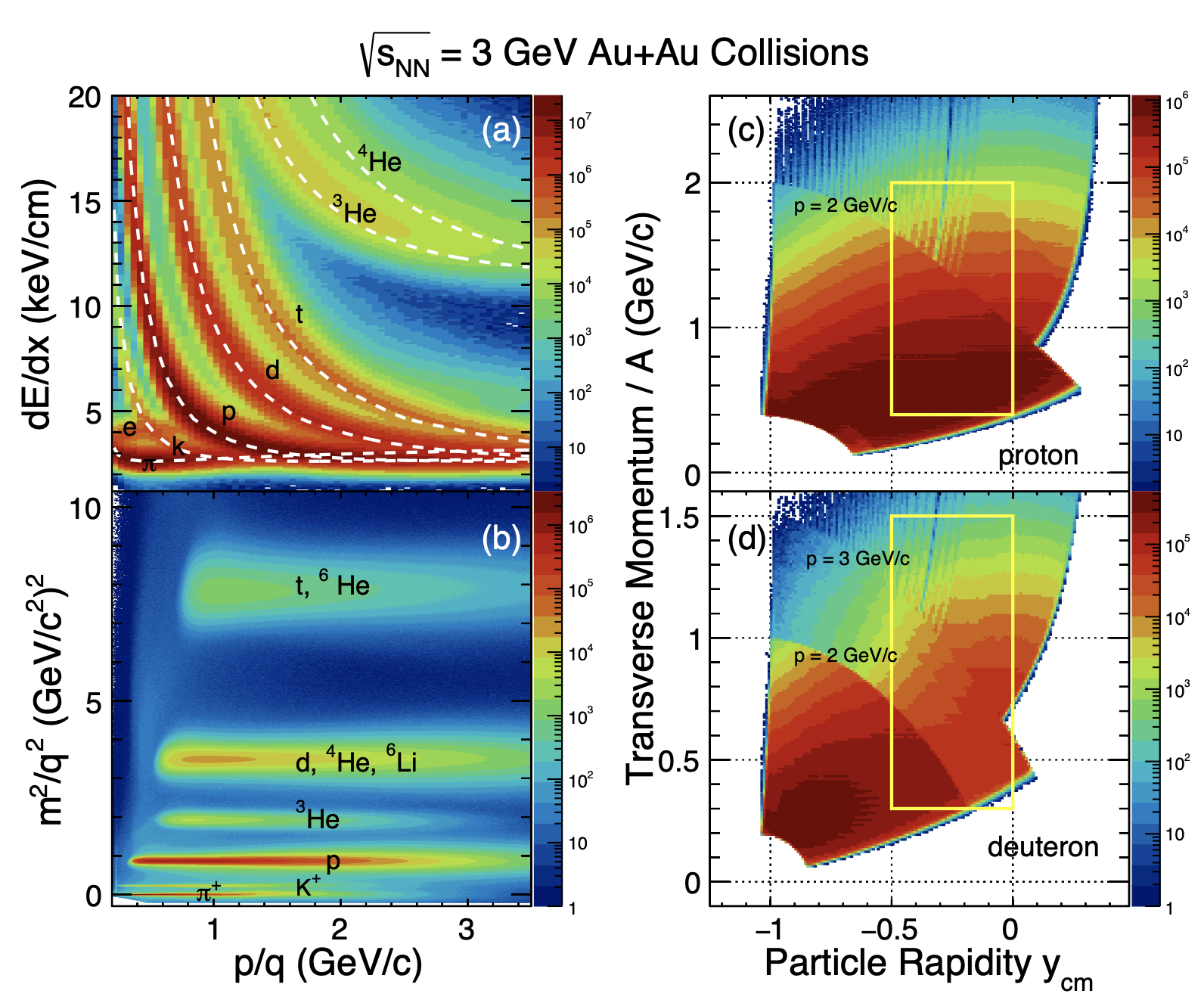}
    \caption{(a) $\langle$$dE/dx$$\rangle$ of charged tracks versus particle rigidity in Au+Au collisions at {\sNN} = 3\,GeV. The dashed lines are Bichsel theoretical curves for the corresponding particle species as labeled. (b) Particle $m^{2}/q^{2}$ versus rigidity. The bands correspond to $\pi^{+}$, $K^{+}$, $p$, $d$, $t$, $^{3}$He and $^{4}$He as labeled. Similar values of $m^{2}/q^{2}$ are found for ($d$, $^{4}$He and $^{6}$Li) and ($t$ and $^{6}$He), respectively.
    (c) and (d) Atomic mass number normalized transverse momentum ($p_{T}/A$) versus rapidity in the center-of-momentum frame ($y_{cm}$) acceptances for $p$ and $d$. The bands in the distributions are caused by the momentum dependent requirements of the PID. The yellow boxes represent the selected phase space for correlation function calculation. Target rapidity is at $y_{cm}=-1.05$.}
\label{fig:PID}
\end{figure}

\begin{figure*}
\includegraphics[width=0.95\linewidth]{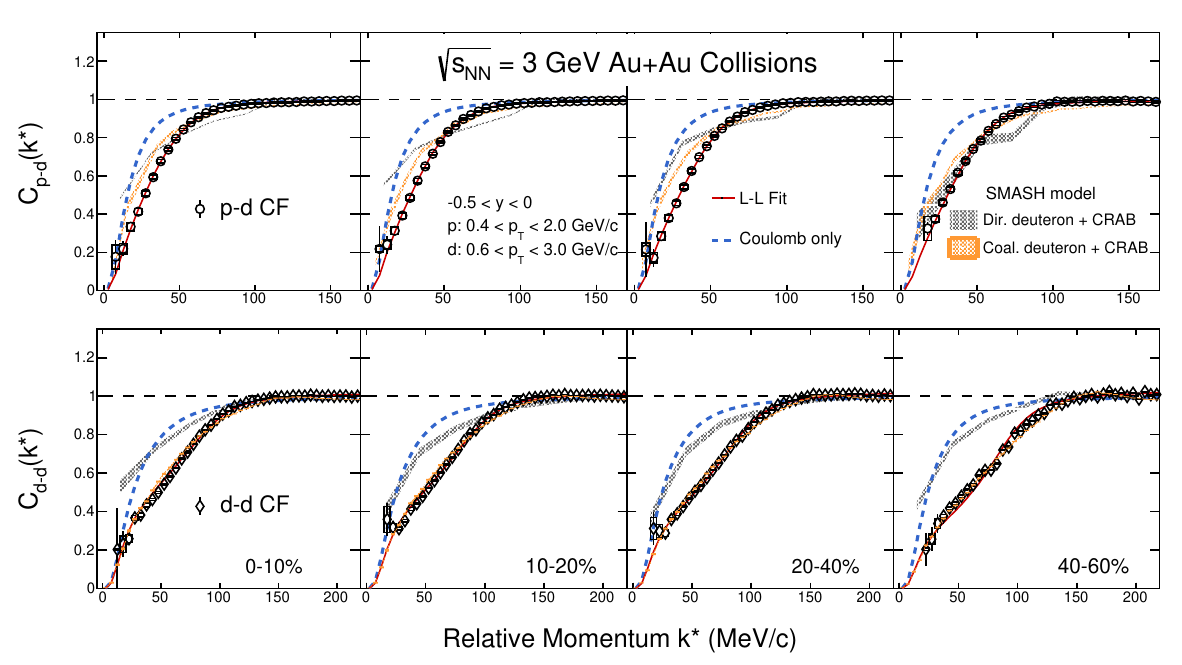}
    \caption{Centrality dependence of the mid-rapidity correlation functions for \pd{}, (top panel) and \dd{}, (bottom panel) displayed as a function of the relative momenta. Statistical and systematic uncertainties from the measurements are shown as bars and boxes, respectively. The results of the Lednick\'y-Lyuboshits (LL) fits are shown as the red-lines. Orange bands represent the calculations from the SMASH model with coalescence procedure for the formation of deuterons plus CRAB afterburner while gray bands show the model calculation with directly produced deuterons plus CRAB. Blue dashed lines are the results with Coulomb interactions only.}\label{fig:CFs}\end{figure*}

Charged-track trajectories are reconstructed from the measured space points information in the TPC. In order to select the primary tracks, each track is required to have a distance of closest approach (DCA) to the event vertex of less than 3 cm. To ensure the quality of reconstructed tracks, each track is required to have at least 15 measured points in the TPC and to have at least 52\% of the possible maximum possible points for its particular geometry. The charged-particle identification is accomplished using the specific energy loss $dE/dx$ measured in the TPC and the reconstructed rigidity. 
Figure~\ref{fig:PID} (a) shows the energy loss $dE/dx$ distribution of protons and deuterons as a function of rigidity. For momentum less than 2 GeV/$c$, the proton and deuteron bands are sufficiently separated that particle identification (PID) can be done cleanly using only the TPC. In order to improve purity for momentum above 2 GeV/$c$, an additional mass-squared cut is performed using PID information from the TOF. Figure~\ref{fig:PID} (b) shows $m^2/q^{2}$ versus momentum with TOF PID. 
Figures~\ref{fig:PID} (c) and (d) show the momentum-space acceptance of selected protons and deuterons as a function of $p_T$ and rapidity ($y_{cm}$) in the center-of-mass frame. The target is at $y_{cm}=-1.05$ and the sign of $y_{cm}$ is chosen such that the beam travels in the positive $y_{cm}$ direction. The $p_{T}-y_{cm}$ acceptance windows used for this analysis are indicated in Figure~\ref{fig:PID} (c,d).

\section{Two-Particle Correlation Function}
The method used to investigate the \pd{} and \dd{} interaction relies on the particle pair correlations measured as a function of $k^*$, defined as the reduced relative momentum of the pair ($k^{*}=|\vec{p}_{1}-\vec{p}_{2}|/2$) in the pair rest frame~\cite{Lednicky:2005af,ALICE:2018ysd,STAR:2015kha}, where $\vec{p}_{1}$ and $\vec{p}_{2}$ are the momenta of the particles. The experimental correlation function is defined as: 
\begin{equation}    
C_{\mathrm{exp}}(k^{*})=\mathcal{N}\frac{N_{\mathrm{same}}(k^{*})}{N_{\mathrm{mixed}}(k^{*})},
\end{equation}
where $N_{\mathrm{same}}(k^{*})$ is the measured distribution of pairs with both particles coming from the same event, $N_{\mathrm{mixed}}(k^{*})$ is the reference distribution generated from mixed events, and $\mathcal{N}$ is a normalization parameter. 
The denominator, $N_{\mathrm{mixed}}(k^{*})$, is obtained by mixing particles from different events that have approximately the same centrality interval (divided into 7 bins: 0-5\%, 5-10\%, 10-20\%, ..., 50-60\%) and vertex position interval along the $z$-direction (a single bin, [198, 202] cm, due to the narrow $V_z$ distribution in FXT). 
The normalization parameter $\mathcal{N}$ is chosen such that the mean value of the correlation function equals unity for 300 $<$ $k^{*}$ $<$ 500 MeV/$c$. Identical single-particle cuts are applied in both same and mixed events. 
The track splitting (one single particle reconstructed as two tracks) and track merging (two particles with similar momenta reconstructed as one track) effects are removed following a standard method used in STAR~\cite{STAR:2004qya}.
The efficiency and acceptance effects cancel in the $N_{\mathrm{same}}(k^{*})/N_{\mathrm{mixed}}(k^{*})$ ratio. The effect of momentum resolution on the correlation functions has been investigated using simulated tracks with known momenta, embedded into real events. More details can be found in ~\cite{STAR:2004qya}. The impact of momentum resolution on correlation functions is found to be less than 1\%. 
Knockout particles from interactions of high-energy particles with detector materials or the beam pipe may contribute to background contamination. A full GEANT simulation of the STAR detector in Au+Au collisions at {\sNN} = 3\,GeV was conducted ~\cite{STAR:2023uxk} and it is found that knockout-particles constitute less than 2\% of the background contamination in the measured acceptance region. Therefore, no knockout correction was applied.

When studying two-particle correlations, it is essential to account for feed-down contributions and PID impurities. In this analysis, the purity of protons and deuterons exceeds 98\%, resulting in a minimal contribution from misidentification. According to the data-driven study~\cite{STAR:2023uxk}, the proton feed-down fraction from weak decays is less than 1\% in the 3 GeV Au+Au collisions, and no feed-down contribution is accounted for deuterons. Given these considerations, corrections to the raw correlation functions were applied according to the following expression:

\begin{equation}
    C_{\mathrm{gen}}(k^*) = \frac{C_{\mathrm{exp}}(k^*) -1 }{PP(k^*)} + 1
\end{equation}
where $PP(k^*)$ is the pair purity.

The Lednick\'y-Lyuboshits (LL) approach~\cite{Lednicky:1981su,Morita:2019rph} is used to parameterize the measured \pd{} and \dd{} correlation function.
Theoretically, the correlation function can be expressed as
\begin{equation}
    C\left(k^{*}\right)=\int S(r)\left|\Psi\left(\overrightarrow{k^{*}}, \vec{r}\right)\right|^{2} \mathrm{~d}^{3} r
\label{eq:KooninPratt}
\end{equation}
where $r$ is the relative distance of the particles that make up the pair of interest, 
$S(r)$ is the incoherent source for particle emission, and $\Psi\left(\overrightarrow{k^{*}}, \vec{r}\right)$ is the wave function describing the relative motion of the pair. 
The wave function for a system of two charged point-like particles is expressed as the sum of a free wave and a scattered spherical wave modified by the scattering amplitude $f_{c}(k^{*})$~\cite{Lednicky:1981su}:
\begin{equation}
\begin{split}
	\Psi\left(\vec{k}^{*}, \vec{r}^{*}\right)=e^{i \delta_{c}} \sqrt{A_{c}(\eta)}\bigg[ e^{-i \vec{k}^{*} \cdot \vec{r}^{*}} F(-i \eta, 1, i \xi) \\ 
	+f^{S}_{c}\left(k^{*}\right) \frac{\tilde{G}(\rho, \eta)}{r *} \bigg]
\end{split}
\end{equation}
where $e^{i\delta_c}=\Gamma(1+i\eta)$ is the Coulomb phase shift factor, $\eta=(k^{*}a_{c})^{-1}$, $a_c$ is the Bohr radius for particle pairs, $\xi=\rho(1+cos(\theta^*))$, $\theta$ is the angle between $\vec{k}^{*}$ and $\vec{r}^{*}$, $\rho=k^{*}r^{*}$, $A_{c}(\eta)=2\pi\eta(e^{2\pi\eta}-1)^{-1}$ is the Coulomb penetration factor. The $F(-i \eta, 1, i \xi)$ is a confluent hypergeometric function and $\tilde{G}(\rho,\eta)=\sqrt{A_{c}}(G_{0}+iF_{0})$ is a combination of the regular ($F_0$) and singular ($G_0$) s-wave Coulomb functions. The scattering amplitude $f^{S}_{c}(k^{*})$, which includes Coulomb interaction, is given by:
\begin{equation}
    f_{c}^{S}\left(k^{*}\right)=\bigg[ \frac{1}{f_{0}^{S}}+\frac{1}{2} d_{0}^{S} k^{* 2}-\frac{2}{a_{c}} h(\eta)-i k^{*} A_{c}(\eta)\bigg]^{-1}
\end{equation}
\begin{equation}
       h(\eta)=\eta^{2} \sum_{n=1}^{\infty} \big[ n(n^{2}+\eta^{2})\big] ^{-1}-C-\ln |\eta|
\end{equation}
(here $C$ = 0.5772 is the Euler constant), 
$f_{0}^{S}$ is the scattering length and $d_{0}^{S}$ is the effective range for a given total spin $S$. For \pd{} interactions, two possible spin configurations are considered,  $S$ = 1/2 and $S$ = 3/2, for a doublet and quartet state, respectively. For \dd{} interactions, one considers $S$ = 0 and $S$ = 2, for a singlet and quintet state. 
For the \dd{} pairs, the S = 1 triplet state is quantum mechanically not allowed after the symmetrized the wave function. 
The different spin configurations cannot be distinguished in these measurements, so spin-averaged results are presented. There is no analytical form of the LL approach that accounts for the Coulomb potential. Therefore, we use numerical integration of Eq.~\ref{eq:KooninPratt} to compute the correlation function. Throughout this paper the standard sign convention is adopted, where, a positive $f_0$ indicates an attractive interaction in a baryon-baryon system, and a negative sign represents a repulsive potential or the presence of a bound state. 
More detailed discussions about the LL approach can be found in Refs.~\cite{Lednicky:1981su,Morita:2019rph,STAR:2015kha}. 
Note the following caveats: (i) The size of deuteron is not considered in our treatment and it is treated as a point-like particle; (ii) The source $S(r)$ is assumed to be a Gaussian type in the fit and it is noted as $R_G$ in the following discussions; (iii) Only s-wave is considered in the wave function~\cite{Lednicky:1981su}; and (iv) as discussed in ~\cite{Lisa:2005dd}, the LL model relies on the smoothness assumption. These effects may become more significant in small colliding systems, such as p+p collisions, however, the relatively large source size observed in heavy-ion collisions justifies the application of the LL approach ~\cite{Lednicky:1981su}. Additionally, the interference between different spins is not considered in this analysis.

\section{Results and Discussion}
The centrality dependence of the mid-rapidity correlation functions for \pd{}, (top panel) and \dd{}, (bottom panel) are displayed as a function of the relative momentum in Fig.~\ref{fig:CFs}. The systematic uncertainties of correlation functions are obtained by varying single-particle selection criteria for protons and deuterons, feed-down contribution to protons, pair selection and normalization range. The resulting uncertainties on the correlation functions are added in quadrature. 
A significant suppression below unity is observed at low relative momentum of the \pd{} and \dd{} correlation functions, indicating an overall repulsive interaction~\cite{ALICE:2023bny}.
Note that in this analysis, the effect of the 3rd-body Coulomb interactions on the final correlation functions is not discussed.

The measured correlation functions are compared to those calculated using the "Simulating Many Accelerated Strongly interaction Hadrons" (SMASH) model~\cite{Weil:2016zrk} under cascade mode. The SMASH is a newly developed hadronic transport model especially suitable for studying the dynamical evolution of heavy-ion collisions at high baryon density~\cite{Weil:2016zrk}. Key features of nuclear collision dynamics including initial condition, baryon stopping~\cite{Schafer:2021csj,Mohs:2019iee}, resonances production and detail balance are implemented in the model. The model offers full phase information of particles at kinetic freeze-out, recorded at the last scattering time, allowing apply experimental cuts which is specially important for realistically determination of collision centrality, the correlation analysis as well as light nuclei production~\cite{Staudenmaier:2021lrg, Mohs:2020awg, STAR:2023uxk}. 
Within the SMASH framework, we consider two distinct production mechanisms for deuterons. 
In the first approach, deuterons were produced through nucleon coalescence, with their formation probability determined by the Wigner function, which takes into account the relative momentum and spatial coordinates of protons and neutrons~\cite{Zhao:2018lyf}.
The second mechanism accounts for directly-produced deuterons through hadron scattering processes (p+n+$\pi\leftrightarrow$ d+$\pi$) in a pion-induced reaction under default setting from SMASH~\cite{Oliinychenko:2018ugs}. This process seems to work for deuteron production at the LHC energies within the experimental uncertainties~\cite{Oliinychenko:2020ply,Oliinychenko:2020znl}.
The calculations with the coalescence mechanism provide a better description of the \dd{} correlation functions, accurately capturing the observed correlations. 
For the \pd{} correlation, the model does not provide a perfect description possibly due to the lack of higher-order wave functions~\cite{Torres-Rincon:2024znb,Rzesa:2024oqp}.	
However in comparison, the directly-produced mechanism tends to underestimate the strength of the correlations, especially for the \dd{} pairs. 
These correlation functions provide further evidence supporting the notion that deuterons are primarily created through the coalescence mechanism, as indicated by the previous measurements of collectivity and yields of light nuclei~\cite{STAR:2021ozh,PhysRevLett.130.202301,Oh:2009gx,Steinheimer:2012tb} in high-energy nuclear collisions. Note that the SMASH model itself does not incorporate the effect of quantum statistics or the final-state interaction (the Coulomb potential and strong interaction) after kinetic freeze-out, hence these are calculated by the "Correlation After Burner (CRAB)~\cite{CRAB}", in which the strong interaction potentials for the \pd{} and \dd{} are adopted from previous studies~\cite{Jennings:1985km}.

\begin{figure}[htbp]
\centering
\includegraphics[width=0.96\linewidth]{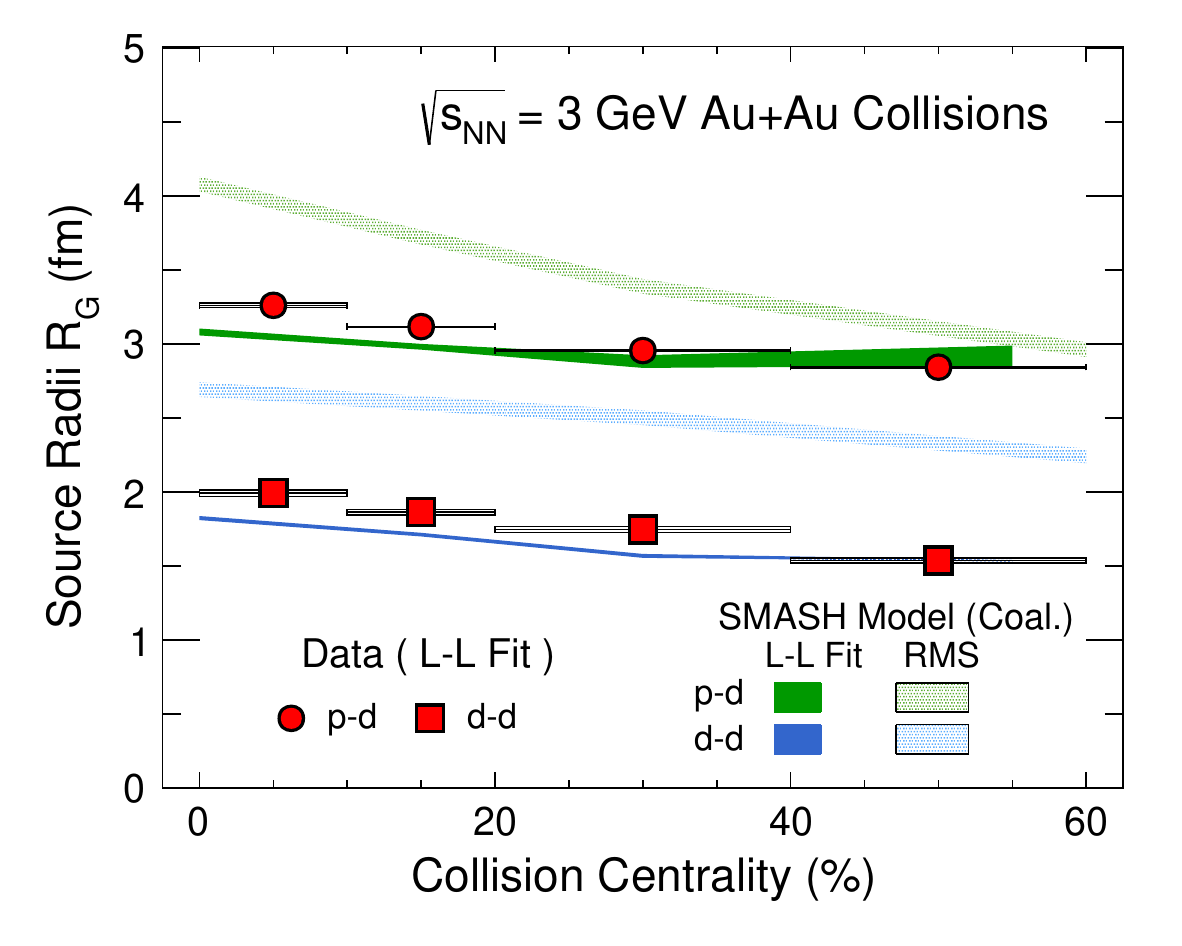}
	\caption{Collision centrality dependence of the source radius parameter extracted from \pd{},(circles) and \dd{},(squares) correlation functions in {\sNN} = 3\,GeV Au+Au collisions. Statistical and systematic uncertainties are all smaller than the size of the symbols. The values of the Gaussian source radius from SMASH model are shown as green and blue bands, for \pd{} and \dd{} pairs, respectively. The shadow bands represent the RMS values calculated from SMASH model.}\label{fig:radii}
\end{figure}

\begin{table*}[!ht]
    \centering
    \caption{The extracted source size $R_G$ parameters in \pd{} and \dd{} pairs with individual fit and simultaneous fit. The errors represent statistical uncertainites from fitting.
}
    \begin{tabular}{c|c|c|c|c}
    \hline\hline
    \multicolumn{5}{c}{\pd{} pair} \\ \hline
    Method    & $R_G$ (0-10\%)  & $R_G$ (10-20\%)  & $R_G$ (20-40\%)  & $R_G$ (40-60\%) \\ \hline
    Indi. Fit & 3.22 $\pm$ 0.04 & 3.11 $\pm$ 0.04 & 2.94 $\pm$ 0.03 & 2.85 $\pm$ 0.05 \\
    Simu. Fit & 3.26 $\pm$ 0.02 & 3.12 $\pm$ 0.02 & 2.95 $\pm$ 0.02 & 2.84 $\pm$ 0.03 \\ \hline\hline
    \multicolumn{5}{c}{\dd{} pair} \\ \hline
    Method    & $R_G$ (0-10\%)  & $R_G$ (10-20\%)  & $R_G$ (20-40\%)  & $R_G$ (40-60\%) \\ \hline
    Indi. Fit & 1.99 $\pm$ 0.02 & 1.87 $\pm$ 0.02 & 1.75 $\pm$ 0.02 & 1.57 $\pm$ 0.03 \\
    Simu. Fit & 1.98 $\pm$ 0.01 & 1.86 $\pm$ 0.01 & 1.75 $\pm$ 0.01 & 1.54 $\pm$ 0.01 \\ \hline\hline
    \end{tabular}
    \label{tab:table_RG}
\end{table*}

The source radius and interaction parameters are extracted by fitting the \pd{} and \dd{} correlation functions with the LL model over the range $k^{*} <$ 200 MeV/c. 
The fitting is performed individually to each centrality ($R_G$, $f_0$ and $d_0$ for each centrality) and simultaneously in four centrality bins to the data (one $R_{G}^{data}$ for each centrality and common $f_0$ and $d_0$). 
To evaluate the systematic uncertainties associated with the fitting procedure, different fitting ranges are considered. 
The extracted $R_{G}^{data}$ for the \pd{} and \dd{} pairs in different centralities, obtained from the LL fit, are listed in Table~\ref{tab:table_RG}. These values were derived through individual fits for each centrality as well as a simultaneous fit across all four centralities. The $R_{G}^{data}$ results from the simultaneous fit are presented in Figure~\ref{fig:radii}.
It is observed that the $R_{G}$ of both \pd{} and \dd{} pairs exhibit a monotonic decrease from central to peripheral collisions. 
Additionally, the $R_{G}^{data}$ of \pd{} is consistently larger than that of \dd{} in all centrality classes. Based on the observation of $m_T$-scaling ~\cite{STAR:2020dav,ALICE:2020ibs,ALICE:2015hvw}, the overall larger $\langle m_{T} \rangle$ of \dd{} pairs  (2.05 $\pm$ 0.01 GeV/c) compared to that of the \pd{} pairs (1.57 $\pm$ 0.01 GeV/c) might explain the observed difference in source radii. A similar phenomenon in \pd{} and \dd{} correlation analyses was also observed in $^{40}$Ar+$^{58}$Ni collisions at 77 MeV/u~\cite{Wosinska:2007rci}. 
As shown in Fig.~\ref{fig:CFs}, the model calculations, using the coalescence for deuteron production plus CRAB for both \pd{} and \dd{} correlation functions, are compatible with the experimental data. The resulting radii $R_{G}^{SMASH}$ from the same LL fit closely match to the data.
As one can see in Figure~\ref{fig:radii}, the root mean square (RMS) values of the source radii directly extracted from SMASH model calculation are larger than the static source radii obtained under the Gaussian assumption. The difference between RMS and $R_G$ can be attributed to the dynamical expansion of the system in nuclear collisions~\cite{STAR:2021yiu,STAR:2005gfr,Li:2008ge}. The resulting source has a time distribution in the model calculation. This leads to a tail in the distance distribution, which in turn cannot be described by a simple Gaussian distribution, making the value of $R_G$ smaller than that of the RMS value. It should be noted that the extracted $R_G$ are comparable to the spatial extent of the deuteron. While the deuteron size may have an impact on the results, the current LL model approximates it as a point-like particle, and the potential effect of its size warrants further investigation. Recently, the ALICE experiment reported a new measurement of the \pd{} correlation function from 13 TeV p+p collisions~\cite{ALICE:2023bny}, where the deuteron size effect was not considered, even in p+p collisions.

\begin{figure*}
    \centering
    \includegraphics[width=0.75\linewidth]{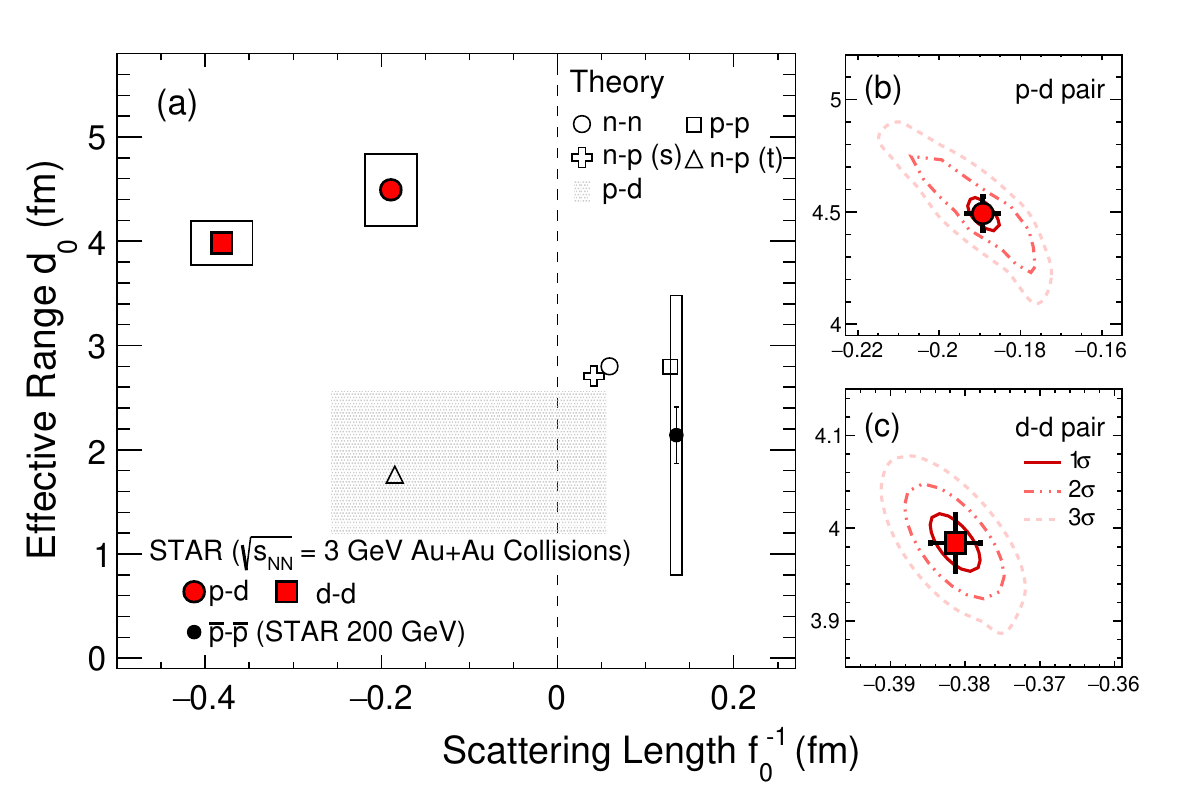}
    \caption{(a) Spin-averaged final state strong interaction parameters: \fzero{} scattering length, and \dzero{}, effective range, extracted from \pd{} (filled circle) and \dd{} (filled square) correlation functions. The statistical uncertainties are smaller than the marker size. Open boxes represent the systematic uncertainties. Data of $\Bar{p}$-$\Bar{p}$ correlation function in 200\,GeV Au+Au collisions~\cite{STAR:2015kha} are shown as the solid black point. The interaction parameters from $n$-$n$, $p$-$p$, $n$-$p$ singlet ($s$), $n$-$p$ triplet ($t$) states~\cite{Mathelitsch:1984hq,Slaus:1989tsq} and $p$-$d$~\cite{Black:1999duc,Huttel:1983wkj,VanOers:1967oww,Kievsky:1997jd,Arvieux:1974fma} are shown as open symbols and hatched area. The 1-3$\sigma$ fitting contours are shown in the right plots (b,c). }
    \label{fig:f0d0}
\end{figure*}

\begin{table*}[!ht]
    \centering
    \caption{The extracted scattering length $f_0$ and effective range $d_0$ parameters from \pd{} and \dd{} correlation functions with individual fit and simultaneous fit. The errors represent statistical uncertainites from fitting.}
    \begin{tabular}{c|c|c|c|c|c}
    \hline\hline
    \multicolumn{6}{c}{\pd{} pair} \\ \hline
	Parameters (fm) & Fit $C_{pd}$(0-10\%) & Fit $C_{pd}$(10-20\%) & Fit $C_{pd}$(20-40\%) & Fit $C_{pd}$(40-60\%) & Simu. Fit \\ \hline
	$f_{0}$     & -5.45 $\pm$ 0.31 & -5.04 $\pm$ 0.25 & -5.53 $\pm$ 0.25 & -5.14 $\pm$ 0.24 &  -5.28 $\pm$ 0.11 \\ \hline
	$d_{0}$     &  4.30 $\pm$ 0.15 &  4.51 $\pm$ 0.15 &  4.46 $\pm$ 0.16 &  4.48 $\pm$ 0.21 &  -4.49 $\pm$ 0.08 \\ \hline
	\multicolumn{6}{c}{\dd{} pair} \\ \hline
	Parameters (fm) & Fit $C_{dd}$(0-10\%) & Fit $C_{dd}$(10-20\%) & Fit $C_{dd}$(20-40\%) & Fit $C_{dd}$(40-60\%) & Simu. Fit \\ \hline
	$f_{0}$         & -2.74 $\pm$ 0.04  & -2.58 $\pm$ 0.04 & -2.57 $\pm$ 0.03 & -2.63 $\pm$ 0.05 &  -2.62 $\pm$ 0.02 \\ \hline
	$d_{0}$		    & 3.88 $\pm$ 0.08 & 3.67 $\pm$ 0.06  & 3.98 $\pm$ 0.04 & 3.90 $\pm$ 0.08 &  -3.99 $\pm$ 0.03 \\ \hline \hline
    \end{tabular}
    \label{tab:table_f0d0}
\end{table*}

The extracted spin-averaged scattering length \fzero{} and effective range \dzero{} obtained from the model fit are listed in Table.~\ref{tab:table_f0d0}, including individual fits for each centrality as well as a simultaneous fit for all four centralities. The results of the simultaneous fit results are shown in Figure ~\ref{fig:f0d0}. 
It is observed that the strong interaction parameters extracted from different centralities are in good agreement within the uncertainties.
For comparison, the interaction parameters for $n$-$p$, $p$-$p$, $n$-$p$ singlet and triplet states ~\cite{Mathelitsch:1984hq,Slaus:1989tsq} as well as $\bar{p}$-$\bar{p}$ measured by STAR~\cite{STAR:2015kha} are also shown in Fig.~\ref{fig:f0d0}. In contrast to the data obtained from $\bar{p}$-$\bar{p}$ correlations and the majority of model predictions for nucleon-nucleon interactions, it is remarkable that the values of the spin-averaged $f_0$ are negative for both \pd{} and \dd{} interactions.
For \pd{} interactions, theoretical predictions ~\cite{Black:1999duc,Huttel:1983wkj,VanOers:1967oww,Kievsky:1997jd,Arvieux:1974fma} suggest that the negative sign of the scattering parameters corresponds to a repulsive interaction in the quartet state and a $^{3}\mathrm{He}$ bound state emerges with a anomalously modified standard effective range expansion (see Eq.(2) in ~\cite{Kievsky:1997jd}). The predicted $1/f_{0}$ of \pd{} interaction spans a range from -0.26 fm to 0.05 fm (hatched area in Fig.~\ref{fig:f0d0}). The experimental measured $f_0$ of \pd{} interactions from the femtoscopy method is in good agreement with the \pd{} interaction parameters from theoretical calculations~\cite{Black:1999duc,Huttel:1983wkj,VanOers:1967oww,Kievsky:1997jd,Arvieux:1974fma}. This consistency supports the feasibility of extracting interaction parameters using the two-particle correlation technique.
From the measurment of the \pd{} correlation function in 13 TeV p+p collisions~\cite{ALICE:2023bny}, in contrast to the heavy-ion results presented here, that analysis indicates that a three-body interaction~\cite{Viviani:2023kxw} seems to be required in order to reproduce the measured \pd{} correlation function in the elementary collisions.
For \dd{} interactions, the negative $f_0$ values can be interpreted similarly, with a repulsive interaction observed in the quintet spin state, alongside the emergence of a $^{4}\mathrm{He}$ bound state. The \dd{} correlation functions and the extracted final state interaction parameters presented here are the first such experimental results from high-energy nuclear collisions.

\section{Summary}
In summary, we have reported the measurements of mid-rapidity light nuclei correlation functions for \pd{} and \dd{} pairs in Au+Au collisions at {\sNN} = 3\,GeV measured by the STAR experiment at RHIC.
The measured correlation functions are reasonably described by the transport model SMASH with a coalescence approach for deuteron formation and afterburner calculations for the correlation functions.
Simultaneous LL fits to the correlation functions with different collision centrality are done to both \pd{} and \dd{} pairs.
The extracted Gaussian equivalent static source radii, $R_{G}^{data}$, display a decreasing trend from central to peripheral collisions. 
Larger values of the radii are observed from the \pd{} pairs compared to those of \dd{} pairs in all centrality bins.
The values of the fitted $R_{G}^{data}$ are found to be consistently smaller than the RMS values extracted from the model calculations implying time-dependent expansion. For the first time, the strong interaction parameters of \pd{} and \dd{} pairs are extracted from heavy-ion collisions. 
The extracted scattering lengths (spin-averaged), $f_0$, are found to be negative, which is consistent with the combination of the repulsive interactions along with the presence of bound state of light nuclei ($^{3}\mathrm{He}$ and $^{4}\mathrm{He}$), respectively. 
Within uncertainties, the measured parameters for the \pd{} interaction are consistent with results of low-energy scattering experiment and model calculations~\cite{Black:1999duc,Huttel:1983wkj,VanOers:1967oww,Kievsky:1997jd,Arvieux:1974fma}. In addition, results from the model calculations imply that coalescence is the dominant process for deuteron formation in the high-energy nuclear collisions.
These systematic measurements of correlation functions and interaction parameters provide valuable insights into the production mechanism of light nuclei and many-body interactions.

\section*{Acknowledgements}
We thank Drs. B. Dönigus, V. Koch, Y. Kamiya, A. Ohnishi, S. Pratt, A. Schwenk for insightful discussions. We thank the RHIC Operations Group and SDCC at BNL, the NERSC Center at LBNL, and the Open Science Grid consortium for providing resources and support.  This work was supported in part by the Office of Nuclear Physics within the U.S. DOE Office of Science, the U.S. National Science Foundation, National Natural Science Foundation of China, Chinese Academy of Science, the Ministry of Science and Technology of China and the Chinese Ministry of Education, NSTC Taipei, the National Research Foundation of Korea, Czech Science Foundation and Ministry of Education, Youth and Sports of the Czech Republic, Hungarian National Research, Development and Innovation Office, New National Excellency Programme of the Hungarian Ministry of Human Capacities, Department of Atomic Energy and Department of Science and Technology of the Government of India, the National Science Centre and WUT ID-UB of Poland, the Ministry of Science, Education and Sports of the Republic of Croatia, German Bundesministerium f\"ur Bildung, Wissenschaft, Forschung and Technologie (BMBF), Helmholtz Association, Ministry of Education, Culture, Sports, Science, and Technology (MEXT), Japan Society for the Promotion of Science (JSPS) and Agencia Nacional de Investigaci\'on y Desarrollo (ANID) of Chile.

\bibliography{draft_PLB_final}
\end{document}